\documentclass{emulateapj}
\usepackage{graphicx}
\usepackage{amsmath}
\def\kms{{\ }{\rm km}\,{\rm s}^{-1}}

\def\ltsima{$\; \buildrel < \over \sim \;$}
\def\simlt{\lower.5ex\hbox{\ltsima}}

\newcommand{\be}{\begin{equation}}
\newcommand{\ee}{\end{equation}}

\begin{document}

\submitted{The Astrophysical Journal Letters, accepted}

\slugcomment{{\it The Astrophysical Journal Letters, accepted}}

\shortauthors{KAZANTZIDIS, ABADI, \& NAVARRO}
\shorttitle{Effects of Galaxy Disks on Halo Shapes} 

\title{The Sphericalization of Dark Matter Halos by Galaxy Disks}

\author{Stelios Kazantzidis\altaffilmark{1}, Mario G.
  Abadi\altaffilmark{2}, and Julio F. Navarro\altaffilmark{3}}

\altaffiltext{1}{Center for Cosmology and Astro-Particle Physics; and
  Department of Physics; and Department of Astronomy, The Ohio State
  University, Columbus, OH 43210, USA; stelios@mps.ohio-state.edu}

\altaffiltext{2}{Instituto de Astronom\'ia Te\'orica y Experimental
  (IATE) Observatorio Astron\'omico de C\'ordoba and CONICET Laprida
  854 X5000BGR C\'ordoba, Argentina; mario@oac.uncor.edu}

\altaffiltext{3}{CIFAR Fellow, Department of Physics and Astronomy, 
  University of Victoria 3800 Finnerty Road, Victoria, BC V8P 5C2,
  Canada; jfn@uvic.ca}

\begin{abstract}
  
  Cosmological simulations indicate that cold dark matter (CDM) halos
  should be triaxial. Validating this theoretical prediction is,
  however, less than straightforward because the assembly of galaxies
  is expected to modify halo shapes and to render them more
  axisymmetric. We use a suite of $N$-body simulations to investigate
  quantitatively the effect of the growth of a central disk galaxy on
  the shape of triaxial dark matter halos. In most circumstances, the
  halo responds to the presence of the disk by becoming more
  spherical. The net effect depends weakly on the timescale of the
  disk assembly but noticeably on the orientation of the disk relative
  to the halo principal axes and it is maximal when the disk symmetry
  axis is aligned with the major axis of the halo. The effect depends
  most sensitively on the overall gravitational importance of the
  disk. Our results indicate that exponential disks whose contribution
  peaks at less than $\sim 50\%$ of their circular velocity are unable
  to modify noticeably the shape of the gravitational potential of
  their surrounding halos.  Many dwarf and low surface brightness
  galaxies are expected to be in this regime, and therefore their
  detailed kinematics could be used to probe halo triaxiality, one of
  the basic predictions of the CDM paradigm. We argue that the complex
  disk kinematics of the dwarf galaxy NGC 2976 might be the reflection
  of a triaxial halo. Such signatures of halo triaxiality should be
  common in galaxies where the luminous component is subdominant.

\end{abstract}


\section{Introduction}
\label{SecIntro}

Cold dark matter (CDM) halos are found to be triaxial in cosmological
$N$-body simulations \citep[e.g.,][]{Frenk_etal88,
  Dubinski_Carlberg91,Jing_Suto02,Bett_etal07}. This finding has
motivated a number of studies designed to constrain halo shapes using
a variety of probes, including the gravitational lensing
of distant galaxies \citep[e.g.,][]{Hoekstra_etal04,
  Mandelbaum_etal06}; the morphology of tidal streams in the Milky Way
(MW) \citep[e.g.,][]{Ibata_etal01, Helmi04, Law_etal09}; the
kinematics of polar ring galaxies \citep[e.g.,][]{Sackett_Sparke90};
the flaring of galactic disks \citep[e.g.,][]{Olling_Merrifield00};
and the X-ray emission from hot gas in galaxies and clusters
\citep[e.g.,][]{Kolokotronis_etal01,Buote_etal02}.

Despite these efforts, a clear picture of either conflict or agreement
with CDM predictions has yet to emerge. This is in part due to the
inherent difficulty of the observational task, but also because most
theoretical predictions rely on simulations where the effects of the
baryonic component of galaxies are neglected. The assembly of a
central galaxy can modify substantially the shape of its surrounding
dark halo \citep[see,
e.g.,][]{Dubinski94,Kazantzidis_etal04a,Abadi_etal10,Tissera_etal10}
by reshaping the box orbits that sustain triaxiality
\citep[e.g.,][]{Debattista_etal08, Valluri_etal10}.  Although there is
reasonable consensus on the qualitative effects of baryons on halo
shapes, there has been little {\it quantitative} work aimed at gauging
the response of realistic triaxial halo models to the changes in the
various parameters that characterize the central galaxy, such as its
mass, size, or the timescale and mode of its assembly.

Here we explore these issues using a series of controlled numerical
simulations where a triaxial halo is evolved under the influence of a
central disk galaxy. Our numerical experiments are similar in nature
to those of \citet{Dubinski94} and \citet{Debattista_etal08}.
However, we extend this work in several respects.  For example, we
focus on the shape of the gravitational potential rather than that of
isodensity contours. The potential is less sensitive to the influence
of substructures, which can induce substantial but transient changes
in the local density whilst affecting little the overall potential.
The gravitational potential is also a quantity of more direct
relevance and applicability to many observational studies of halo
shapes.

Further, we explore systematically several aspects of the growth of
the galaxy that may in principle affect the shape of its surrounding
halo. We consider not only the gravitational importance of the disk,
but also its orientation relative to the halo principal axes, as well
as the timescale and mode of its assembly. Lastly, we use halo
parameters in agreement with the results of cosmological $N$-body
simulations and galaxy parameters which are consistent with observed
scaling laws and span the wide range in surface brightness of observed
galaxy disks.


\begin{figure*}[t]
\centerline{\epsfxsize=5in\epsffile{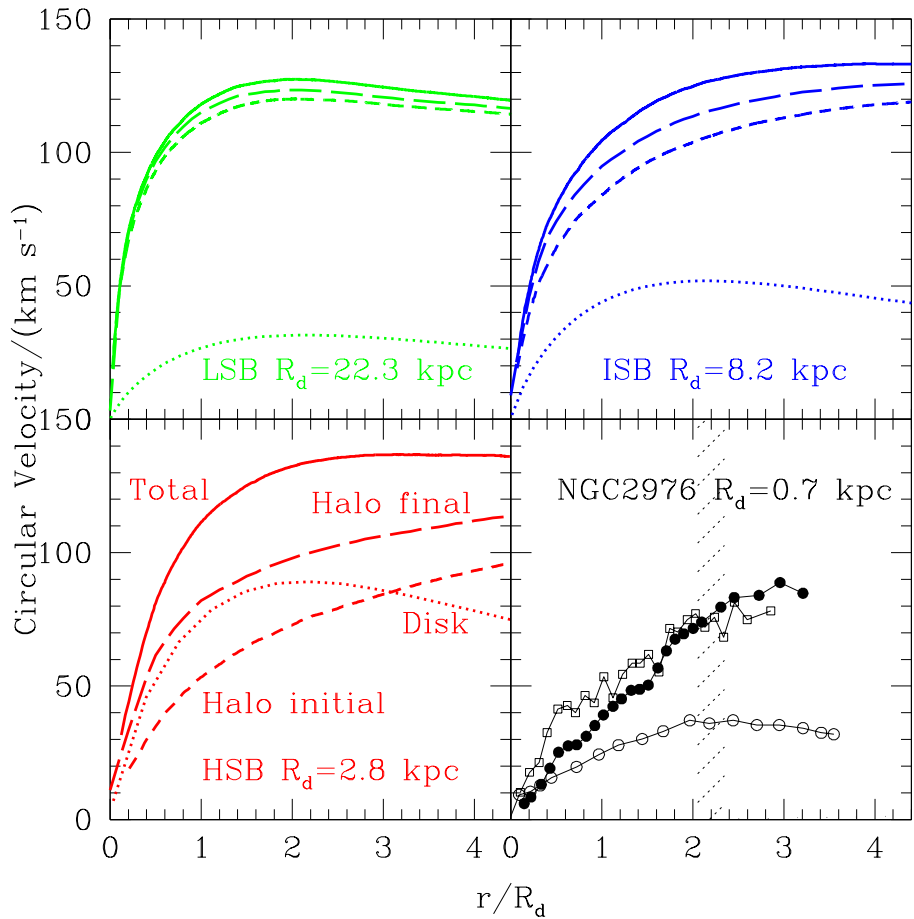}}
\caption{{\it Top and bottom-left panels:} Circular velocity profiles
  of triaxial halo model A. Short and long-dashed lines correspond,
  respectively, to the halo before and after the addition of the disk
  (dotted line).  The disk mass is, in all cases, $M_d \sim 1.3 \times
  10^{10} M_{\odot}$, but its scale length varies: $R_d = 22.3$, $8$,
  and $2.8$~kpc. Galaxy masses are consistent with the baryonic
  TF relation for a rotation speed of $\sim 120 \kms$. The
  disk models span roughly two decades in surface density (or $5$ mag
  in surface brightness) from values typical of LSB galaxies to
  intermediate (ISB) to ``normal'' (HSB) spirals. Radii are given in
  units of $R_d$. The halo circular velocity profile
  peaks at $V_{\rm max} \sim 120 \kms$ at $r_{\rm max} \sim 48$~kpc.
  {\it Bottom-right:} Circular velocity profile (top two curves) and
  disk contribution to the circular velocity (bottom curve) for NGC
  2976. The disk contribution is taken from \citet{Simon_etal03}.
  Circular velocities are taken from \citet{Simon_etal03} (filled
  circles) and \citet{Spekkens_Sellwood07} (open squares).  The shaded
  area indicates the radius, $r= 2.2 R_d$, where the disk
  gravitational importance parameter $\eta=V_d/V_{\rm circ}$ is
  measured. For NGC 2976, $R_d \sim 0.7$~kpc and $\eta \sim 0.5$.
  \label{fig1}}
\end{figure*}


\section{Simulations and Methods}
\label{SecNumSim}

We investigate changes in the shapes of triaxial dark matter (DM)
halos induced by the growth of central disk galaxies modeled as rigid
potentials (see \citealt{Villalobos_etal10} for details). We measure
the shape of the gravitational potential by approximating the
isopotential surfaces by ellipsoids and calculating the principal axes
$a$, $b$, and $c$ \citep[e.g.,][]{Springel_etal04,
  Hayashi_etal07,Abadi_etal10}, where $a$, $b$, and $c$ denote the
major, intermediate, and minor axis, respectively.

The triaxial DM halo models are constructed via successive mergers
\citet{Moore_etal04}.  The initial system is a $75,000$-particle,
isotropic, spherically-symmetric model generated by sampling the
\citet{Hernquist90} distribution function \citep{Kazantzidis_etal04b}.
A nearly prolate remnant results from a head-on collision between two
such Hernquist models; more triaxial configurations are built by
merging this remnant with other Hernquist models on parabolic orbits
of various inclinations relative to the principal axes of the first
remnant.  Each final triaxial halo consists of $300,000$ particles.
After the mergers are complete, we evolve the remnants for several
dynamical times in order to ensure that equilibrium is achieved. The
final configurations have mass distributions that can be well
approximated by NFW profiles \citep{Navarro_etal96} (in the regions
relevant to the present study) and are thus consistent with the
results of cosmological CDM simulations.


\begin{figure*}[t]
\centerline{\epsfxsize=4in\epsffile{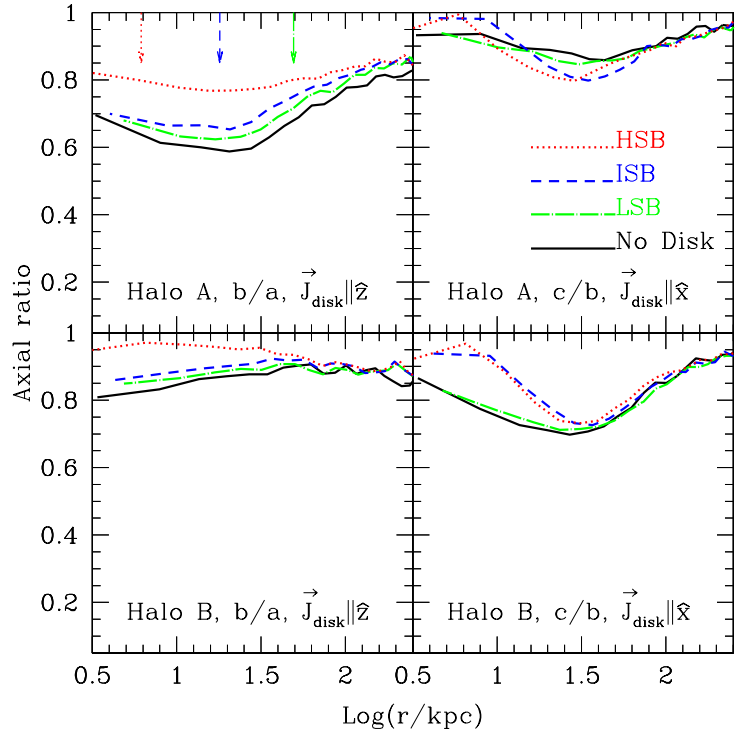}}
\caption{Axial ratios of the {\it halo} gravitational potential
  measured {\it in the plane of the disk} before (solid lines) and
  after the growth of a central galaxy, as a function of radius.
  Arrows indicate the radius ($r=2.2 R_d$) where the
  disk contribution to the circular velocity is maximum.  The {\it
    dot-dashed}, {\it dashed}, and {\it dotted} lines show halo axial
  ratios after the growth of the LSB, ISB, and HSB disk, respectively.
  Panels on the left show the case where the disk plane is
  perpendicular to the {\it minor} ($z$) axis of the halo, while
  panels on the right correspond to the case where the disk plane is
  perpendicular to the {\it major} ($x$) axis of the halo. Results are
  presented for $\tau_d=10$~Gyr but are insensitive to $\tau_d$.
  \label{fig2}}
\end{figure*}


We adopt two different halo models in order to grow central disks. The
first (model A) is a nearly prolate merger remnant, with triaxiality
parameter \citep{Franx_etal91} $T = (a^2 - b^2) / (a^2 - c^2) \sim
0.8$ for most radii. The second (model B) is strongly triaxial; $T
\sim 0.4$ in the inner regions, increasing to $T \sim 0.6$ in the
outskirts. These shapes are fairly typical of those found in
cosmological simulations \citep{Hayashi_etal07}. The spin parameter
$\lambda$ of halos A and B are 0.024 and 0.040, respectively, which
are typical of cosmological halos \citep{Maccio_etal08}. The circular
velocity and axial ratio profiles of these halos are presented in
Figures~\ref{fig1} and \ref{fig2}, respectively.

We scale the parameters of halos A and B to match the mass and
concentration of CDM halos with maximum circular velocity $V_{\rm max}
\sim 120 \kms$. For a halo of virial mass $M_{\rm vir} = 5 \times
10^{11} M_{\odot}$ and concentration $c \sim 9.3$ \citep{Neto_etal07},
the circular velocity profile peaks at $r_{\rm max} \sim 48$~kpc.

Disk galaxies forming in our triaxial halos must have rotation speeds
comparable to $V_{\rm max}$ in order to satisfy simultaneously the
normalization of the Tully-Fisher (TF) relation and the galaxy stellar
mass function \citep[see, e.g.,][]{Croton_etal06,Guo_etal09}.
According to the baryonic TF relation of
\citet{Noordermeer_Verheijen07}, the mass of the luminous component of
galaxies with rotation speeds of order $120 \kms$ is $M_d \sim 1.3
\times 10^{10} M_{\odot}$.  We model these galaxies as exponential
disks with three different values of the radial scale length,
$R_d=2.8, 8.0$, and $22.3$~kpc.  (For the disk vertical structure we
assume an isothermal sheet with scale height $z_d=0.2\,R_d$,
consistent with observations of external galaxies.)  These values span
the range of observed scale lengths for galaxies of this mass, from
the very extended low surface brightness (LSB) disks to intermediate
(ISB) to concentrated high surface brightness (HSB) ``normal''
spirals.

Within the disk radius the circular speed of the disk$+$halo system is
roughly the same for all models, irrespective of the surface
brightness/density of the disk. This is in agreement with
observations, which show that the zero point of the TF relation is
roughly independent of surface brightness \citep{Zwaan_etal95}.


\begin{figure*}
\centerline{\epsfxsize=5in\epsffile{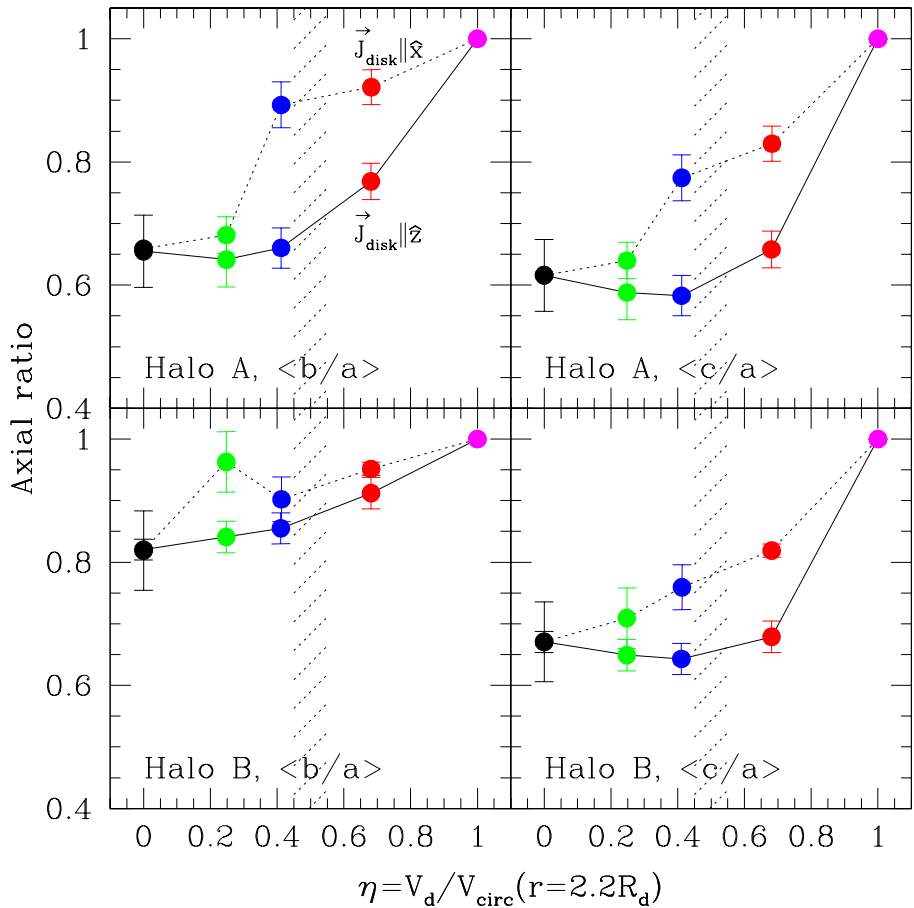}}
\caption{Axial ratios of the inner ($r<30$~kpc) potential, as a function of the disk gravitational 
  importance parameter, $\eta= V_d / V_{\rm circ}$, measured at $r=2.2
  R_d$.  Symbols and error bars indicate the average and the rms shape
  of the halo isopotential contours within $30$~kpc, respectively. The
  leftmost point in each panel ($\eta=0$) corresponds to the halo {\it
    before} adding the disk, while the rightmost point ($\eta=1$)
  corresponds to a fully dominant axisymmetric disk. Intermediate
  values of $\eta$ show results for disks of different surface
  density: LSB, ISB, and HSB, from left to right. Results correspond
  to a disk growth timescale of $\tau_d=10$~Gyr. The orientation of
  the disk plane relative to the principal axes of the halo in each
  panel is indicated in the labels. For given $\eta$, the effect is
  maximal when the disk symmetry axis is aligned with the halo major
  axis. The shaded area highlights the value of $\eta$ of NGC 2976.
\label{fig3}}
\end{figure*}


\begin{figure*}
\centerline{\epsfxsize=5in\epsffile{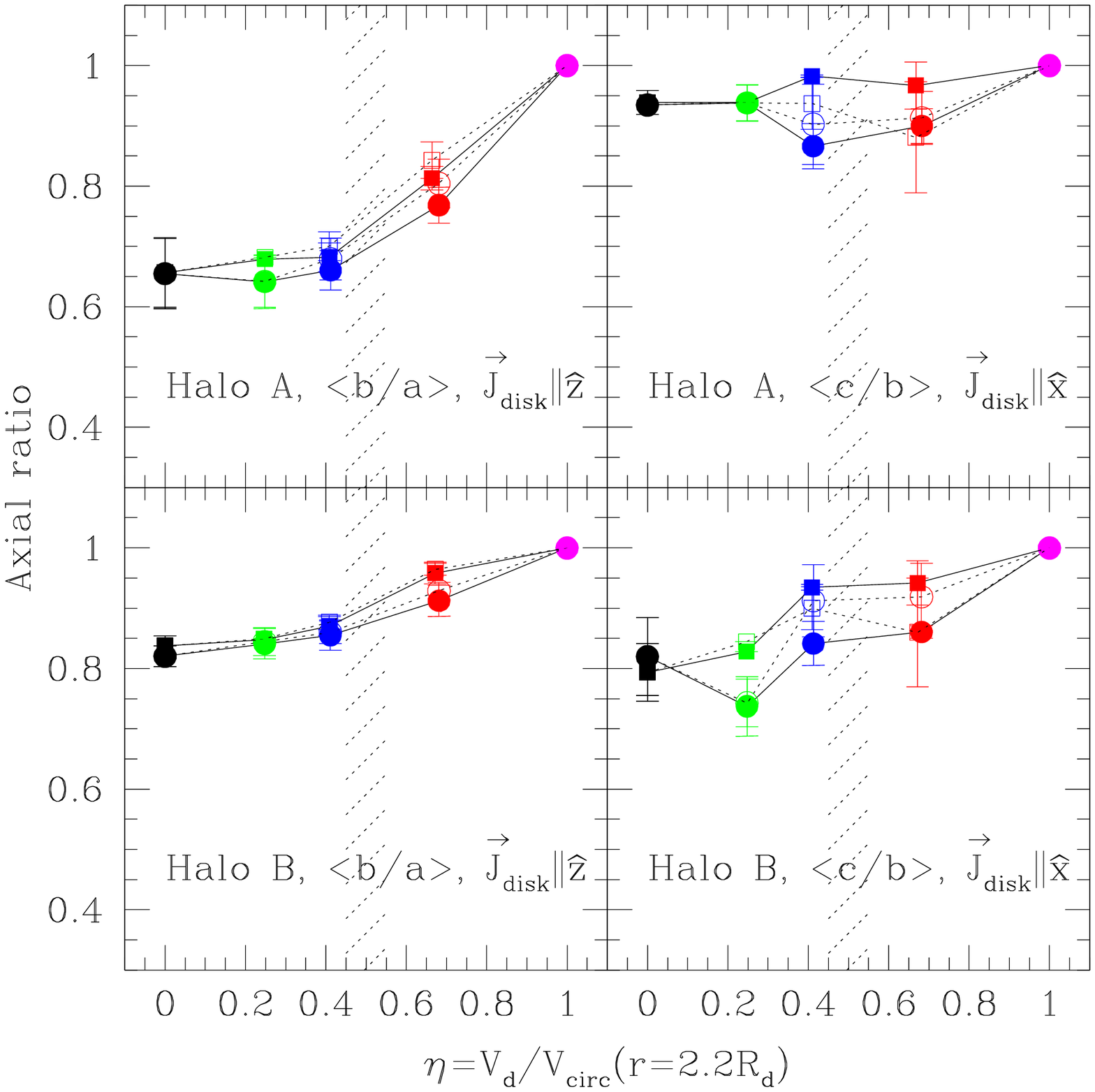}}
\caption{Axial ratios of the inner ($r<30$~kpc) 
  potential, computed {\it in the plane of the disk}, as a function of
  $\eta$. Symbols and error bars are as in Figure~\ref{fig3}.  Filled
  symbols show the shape of the halo isopotential contours; open
  symbols correspond to the total (disk+halo) potential. Circles and squares show 
  cases where the disk growth timescale is $\tau_d$=$0$~Gyr and $\tau_d$=$10$~Gyr,
  respectively.  The orientation of the disk plane relative to the
  principal axes of the halo in each panel is as in Figure~\ref{fig2}.
  \label{fig4}}
\end{figure*}


In addition to surface brightness, we have also explored the response
of the halo to varying the timescale of disk assembly. This is
accomplished by allowing the disk mass to grow linearly over three
different timescales, $\tau_d$=$0$~Gyr, $1$~Gyr, and $10$~Gyr.  The
third choice ensures that the response of the halo is adiabatic,
whereas the first, although unrealistic, provides a useful check of
the sensitivity of our results to $\tau_d$. We have also varied the
mode of disk assembly by simulating cases where the mass of the disk
grows in three discrete steps over $1$~Gyr, instead of linearly. None
of the results we obtain depend on either the choices for $\tau_d$ or
the mode of disk assembly (see also
\citealt{Berentzen_Shlosman06}), so we shall mainly concentrate on
the ``adiabatic'' ($\tau_d=10$~Gyr) experiments where the mass of the
disk grows linearly with time.

Because the precise alignment between the disk angular momentum axis
and the host halo principal axes is still a matter of debate
\citep[e.g.,][]{Faltenbacher_etal05,Zentner_etal05,Bailin_etal05}, the
final parameter we investigated is the orientation of the disk plane
relative to the principal axes of the triaxial halo. For each halo, we
placed the disk plane perpendicular to its major as well as its minor
axis. In total, we performed $24$ simulations of the growth of a
central disk galaxy in each of the triaxial halo models A and B, for a
total of $48$ experiments. After the disk growth is complete, all
simulations are continued for at least a further $1.5$~Gyr in order to
allow for equilibrium to be reached.  In the experiments where the
disk grows in three distinct events, one-third of the disk mass
increases instantaneously at $0.1$, $0.6$, and $1.1$~Gyr. We only
consider the final equilibrium configurations of the halos when
measuring the shape of the potential.

All numerical simulations were carried out with the parallel $N$-body
code PKDGRAV \citep{Stadel01}.  The softening was set to
$\epsilon=0.5$~kpc.  Each disk was modeled with $N=10^5$ static
particles in order to ensure a smooth representation of the disk
potential.

\section{Results} 
\label{SecRes}
 
Figure~\ref{fig2} shows the axial ratios of the halo isopotential
contours measured {\it in the plane of the disk}, for various
experiments. The growth of the disk modifies the halo shape, making it
more axisymmetric. Halo A is nearly prolate so placing the disk plane
perpendicular to the major axis has little effect on the shape of the
2D potential on the disk plane. When the disk plane contains the major
axis the potential is strongly non-axisymmetric, but becomes rounder
after the disk is added. The effect, however, is minor for the case of
the LSB and ISB disk; only the HSB galaxy is able to modify
substantially the halo shape, increasing the inner axial ratio from
$\sim 0.6$ to $\sim 0.8$.

Before adding the disk, halo B is rather triaxial, with $b/a \sim c/b
\sim 0.8$ in the inner regions. In the case of the HSB disk, placing
the disk plane perpendicular to the halo minor axis renders the halo
potential almost perfectly axisymmetric (measured in the plane of the
disk). On the other hand, the LSB and the ISB galaxies are again
barely able to modify the shape of the halo. When the disk plane is
perpendicular to the halo major axis the results are similar, although
in this case the halo response to the ISB and HSB galaxies are
comparable. Except very near the center, the potential remains far
from axisymmetric in all cases. We note that the change in shape in
the case of the HSB disk is noticeable out to $\sim 30$-$50$~kpc, well
outside the region where the disk is gravitationally important.

Figure~\ref{fig3} shows how the 3D halo shape changes in response to
the disk growth. The axial ratios of the inner gravitational potential
are presented as a function of the peak contribution of the disk to
the circular velocity. This is measured by the parameter $\eta = V_d /
V_{\rm circ}$, computed at $r=2.2 R_d$, the radius where the
exponential disk contribution to the circular velocity is maximal.
The shape of the potential is not exactly constant in the inner
regions (see Fig.~\ref{fig2}), and therefore we show the shape of the
potential averaged inside $30$~kpc.  The trends we discuss are robust
to reasonable changes in the averaging procedure.

Results are presented for the ``adiabatic'' ($\tau_d=10$~Gyr) disk
growth and for both orientations of the disk plane relative to the
halo principal axis. $\eta$ varies from $\sim 0.25$ for the LSB galaxy
to $\sim 0.7$ for the HSB model. Figure~\ref{fig3} demonstrates that
in most circumstances the disk growth renders the 3D potential of the
halo more spherical. Because halo accelerations are minimal along its
major axis, whereas the disk gravity is strongest along its symmetry
axis, the effect of a disk on a triaxial halo is maximal when the disk
symmetry axis is aligned with the halo major axis, and minimal in the
opposite case.

Figures~\ref{fig2} and~\ref{fig3} show that the modifications to halo
shapes induced by the disks are relatively minor, except for the HSB
galaxy.  The response of the halo depends principally on the disk
gravitational importance. These results seem insensitive to the
timescale and mode of disk assembly.  Figure~\ref{fig4} shows this
explicitly. This figure shows the axial ratio of the inner potential
as a function of $\eta$ computed {\it in the plane of the disk} and
confirms the conclusions advanced above.  The change in shape is
monotonic with $\eta$ and is relatively minor for the LSB and ISB
models. Only the HSB galaxy is able to modify the halo potential
noticeably. The transition seems to occur roughly at $\eta \approx
0.5$.  Disks that contribute less than about $50\%$ of its circular
velocity should be surrounded by halos that have preserved their
original triaxiality and, therefore, should in general show signatures
of departures from axisymmetry. This is the regime of many LSB and
dwarf galaxies, and it is therefore important to consider the
possibility that their halos may be triaxial when analyzing their
kinematics. \citet{Berentzen_Shlosman06} also concluded that maximal
disks probably reside in nearly axisymmetric halos.

\section{Discussion} 
\label{SecDisc}

The sphericalization of DM halos by central galaxies has important
implications for the interpretation of observational constraints. For
example, a nearly spherical halo is favored by the weak precession of
the Sagittarius stream, a result that would be difficult to understand
in the CDM paradigm unless the MW halo is an outlier in the
distribution of halo shapes
\citep[e.g.,][]{Ibata_etal01,Majewski_etal03}. Indeed, given the
gravitational importance of the MW disk ($\eta \sim 0.8$ for the
canonical values of $V_{\rm rot}=220 \kms$, $R_d=3$~kpc, and
$M_d=5\times 10^{10} M_{\odot}$), the MW halo should have been
significantly affected by the disk.  Nearly spherical halos may also
affect the dynamics of bars and stars in galactic disks
\citep[e.g.,][]{Ideta_Hozumi00,El-Zant_Shlosman02}, and can inhibit
the fueling of nuclear supermassive black holes by suppressing the
large-scale asymmetries responsible for the loss of angular momentum
in the gas component
\citep[e.g.,][]{Merritt_Quinlan98,Callegari_etal10}.

On the other hand, our results suggest that LSB halos should retain
the triaxiality predicted by cosmological simulations. This may help
to explain the rather elongated shapes of the ultra-faint MW
satellites \citep{Martin_etal08} (although these can also be
attributed to tidal effects in the gravitational field of the MW) and
oddities in the kinematics of some dwarf spheroidal galaxies
\citep[see, e.g.,][]{Penarrubia_etal10}. Many dwarf galaxies are also
in the regime $\eta \simlt 0.5$ where they are unable to modify their
surrounding halos. Gaseous disks in such galaxies should exhibit
departures from axisymmetry, unless the halos are either prolate or
oblate and the disk plane coincides with that where the 2D potential
is axisymmetric.

This could indeed be the case in nearly oblate halos, given the
preference of the angular momentum to align with the minor axis
\citep{Bett_etal07}. However, in nearly prolate halos (a more common
occurrence according to $N$-body simulations) disks whose angular
momentum aligns with the minor axis would feel a non-axisymmetric 2D
gravitational potential. Non-circular motions should therefore be
fairly common in the gaseous disks of dwarfs, and they could in
principle be used to gauge the triaxiality of their surrounding halos.

To first order, a gaseous disk in the non-axisymmetric potential of a
triaxial halo would behave just like gas in a barred potential where
the pattern speed of the bar is zero. A subdominant disk in a triaxial
potential would thus exhibit the non-circular dynamical signature of a
(slow) bar but with no obvious bar in the luminous distribution.

Interestingly, there is one system where all these conditions are met.
NGC 2976 is a nearby dwarf spiral galaxy whose baryonic disk is
subdominant, as shown in the bottom-right panel of Figure~\ref{fig1}.
This figure shows that the contribution of the baryonic component
peaks at about one-half of the circular velocity at $r=2.2 R_d$ and
therefore $\eta\approx 0.5$.  \citet{Simon_etal03} show that the
kinematics of the gaseous disk in NGC 2976 is highly complex,
exhibiting large non-circular motions near the center. These,
according to \citet{Spekkens_Sellwood07}, are best understood as the
characteristic kinematic asymmetries imposed by an $m=2$ bar mode in
the gravitational potential \citep[see also][]{Hayashi_Navarro06}. On
the other hand, NGC 2976 has no obvious bar, at least in the optical
\citep[but see][]{Menendez_etal07}, so ascribing the origin of the
non-circular motions to halo triaxiality is clearly tempting.

If this interpretation is correct, then it would be surprising if
other galaxies with subdominant baryonic components did not also show
signs of being embedded in triaxial potentials. Indeed, one may even
argue that the absence of such signatures in a significant fraction of
unbarred LSB and dwarf galaxies would be quite difficult to
accommodate within the standard CDM paradigm.  Definitive conclusions
on these issues require more sophisticated theoretical modeling of the
formation of subdominant disks in triaxial halos. Note that our models
neglect, for example, the response of the disk to the triaxial forcing
of the halo, as well as a realistic accounting of the distribution of
disk orientations relative to the principal axes of the halo.
Nevertheless, our results suggest that a careful search for signatures
of halo triaxiality in a statistically significant sample of dwarf and
LSB galaxies would be warranted. Steps in this direction such as those
taken by \citet{Trachternach_etal09} should certainly be encouraged.

\acknowledgments

The authors would like to thank Josh Simon and Isaac Shlosman for
stimulating discussions.  SK is supported by the Center for Cosmology
and Astro-Particle Physics (CCAPP) at The Ohio State University. MGA
is grateful for the hospitality of CCAPP during the initial stages of
this work. This research was supported by an allocation of computing
time from the Ohio Supercomputer Center (http://www.osc.edu).

\bibliography{ms} 

\begin{thebibliography}{50}
\expandafter\ifx\csname natexlab\endcsname\relax\def\natexlab#1{#1}\fi

\bibitem[{{Abadi} {et~al.}(2010){Abadi}, {Navarro}, {Fardal}, {Babul}, \&
  {Steinmetz}}]{Abadi_etal10}
{Abadi}, M.~G., {Navarro}, J.~F., {Fardal}, M., {Babul}, A., \& {Steinmetz}, M.
  2010, MNRAS, in press (astro-ph/0902.2477)

\bibitem[{{Bailin} {et~al.}(2005){Bailin}, {Kawata}, {Gibson}, {Steinmetz},
  {Navarro}, {Brook}, {Gill}, {Ibata}, {Knebe}, {Lewis}, \&
  {Okamoto}}]{Bailin_etal05}
{Bailin}, J., {Kawata}, D., {Gibson}, B.~K., {Steinmetz}, M., {Navarro}, J.~F.,
  {Brook}, C.~B., {Gill}, S.~P.~D., {Ibata}, R.~A., {Knebe}, A., {Lewis},
  G.~F., \& {Okamoto}, T. 2005, \apjl, 627, L17

\bibitem[{{Berentzen} \& {Shlosman}(2006)}]{Berentzen_Shlosman06}
{Berentzen}, I. \& {Shlosman}, I. 2006, \apj, 648, 807

\bibitem[{{Bett} {et~al.}(2007){Bett}, {Eke}, {Frenk}, {Jenkins}, {Helly}, \&
  {Navarro}}]{Bett_etal07}
{Bett}, P., {Eke}, V., {Frenk}, C.~S., {Jenkins}, A., {Helly}, J., \&
  {Navarro}, J. 2007, \mnras, 376, 215

\bibitem[{{Buote} {et~al.}(2002){Buote}, {Jeltema}, {Canizares}, \&
  {Garmire}}]{Buote_etal02}
{Buote}, D.~A., {Jeltema}, T.~E., {Canizares}, C.~R., \& {Garmire}, G.~P. 2002,
  \apj, 577, 183

\bibitem[{{Callegari} {et~al.}(2010){Callegari}, {Kazantzidis}, {Mayer},
  {Colpi}, {Bellovary}, {Quinn}, \& {Wadsley}}]{Callegari_etal10}
{Callegari}, S., {Kazantzidis}, S., {Mayer}, L., {Colpi}, M., {Bellovary},
  J.~M., {Quinn}, T., \& {Wadsley}, J. 2010, ApJ submitted (astro-ph/1002.1712)

\bibitem[{{Croton} {et~al.}(2006)}]{Croton_etal06}
{Croton}, D.~J. {et~al.} 2006, \mnras, 365, 11

\bibitem[{{Debattista} {et~al.}(2008)}]{Debattista_etal08}
{Debattista}, V.~P. {et~al.} 2008, \apj, 681, 1076

\bibitem[{{Dubinski}(1994)}]{Dubinski94}
{Dubinski}, J. 1994, \apj, 431, 617

\bibitem[{{Dubinski} \& {Carlberg}(1991)}]{Dubinski_Carlberg91}
{Dubinski}, J. \& {Carlberg}, R.~G. 1991, \apj, 378, 496

\bibitem[{{El-Zant} \& {Shlosman}(2002)}]{El-Zant_Shlosman02}
{El-Zant}, A. \& {Shlosman}, I. 2002, \apj, 577, 626

\bibitem[{{Faltenbacher} {et~al.}(2005){Faltenbacher}, {Allgood},
  {Gottl{\"o}ber}, {Yepes}, \& {Hoffman}}]{Faltenbacher_etal05}
{Faltenbacher}, A., {Allgood}, B., {Gottl{\"o}ber}, S., {Yepes}, G., \&
  {Hoffman}, Y. 2005, \mnras, 362, 1099

\bibitem[{{Franx} {et~al.}(1991){Franx}, {Illingworth}, \& {de
  Zeeuw}}]{Franx_etal91}
{Franx}, M., {Illingworth}, G., \& {de Zeeuw}, T. 1991, \apj, 383, 112

\bibitem[{{Frenk} {et~al.}(1988){Frenk}, {White}, {Davis}, \&
  {Efstathiou}}]{Frenk_etal88}
{Frenk}, C.~S., {White}, S.~D.~M., {Davis}, M., \& {Efstathiou}, G. 1988, \apj,
  327, 507

\bibitem[{{Guo} {et~al.}(2010){Guo}, {White}, {Li}, \&
  {Boylan-Kolchin}}]{Guo_etal09}
{Guo}, Q., {White}, S., {Li}, C., \& {Boylan-Kolchin}, M. 2010, \mnras, 404,
  1111

\bibitem[{{Hayashi} \& {Navarro}(2006)}]{Hayashi_Navarro06}
{Hayashi}, E. \& {Navarro}, J.~F. 2006, \mnras, 373, 1117

\bibitem[{{Hayashi} {et~al.}(2007){Hayashi}, {Navarro}, \&
  {Springel}}]{Hayashi_etal07}
{Hayashi}, E., {Navarro}, J.~F., \& {Springel}, V. 2007, \mnras, 377, 50

\bibitem[{{Helmi}(2004)}]{Helmi04}
{Helmi}, A. 2004, \mnras, 351, 643

\bibitem[{{Hernquist}(1990)}]{Hernquist90}
{Hernquist}, L. 1990, \apj, 356, 359

\bibitem[{{Hoekstra} {et~al.}(2004){Hoekstra}, {Yee}, \&
  {Gladders}}]{Hoekstra_etal04}
{Hoekstra}, H., {Yee}, H.~K.~C., \& {Gladders}, M.~D. 2004, \apj, 606, 67

\bibitem[{{Ibata} {et~al.}(2001){Ibata}, {Lewis}, {Irwin}, {Totten}, \&
  {Quinn}}]{Ibata_etal01}
{Ibata}, R., {Lewis}, G.~F., {Irwin}, M., {Totten}, E., \& {Quinn}, T. 2001,
  \apj, 551, 294

\bibitem[{{Ideta} \& {Hozumi}(2000)}]{Ideta_Hozumi00}
{Ideta}, M. \& {Hozumi}, S. 2000, \apjl, 535, L91

\bibitem[{{Jing} \& {Suto}(2002)}]{Jing_Suto02}
{Jing}, Y.~P. \& {Suto}, Y. 2002, \apj, 574, 538

\bibitem[{{Kazantzidis} {et~al.}(2004{\natexlab{a}}){Kazantzidis}, {Kravtsov},
  {Zentner}, {Allgood}, {Nagai}, \& {Moore}}]{Kazantzidis_etal04a}
{Kazantzidis}, S., {Kravtsov}, A.~V., {Zentner}, A.~R., {Allgood}, B., {Nagai},
  D., \& {Moore}, B. 2004{\natexlab{a}}, \apjl, 611, L73

\bibitem[{{Kazantzidis} {et~al.}(2004{\natexlab{b}}){Kazantzidis}, {Magorrian},
  \& {Moore}}]{Kazantzidis_etal04b}
{Kazantzidis}, S., {Magorrian}, J., \& {Moore}, B. 2004{\natexlab{b}}, \apj,
  601, 37

\bibitem[{{Kolokotronis} {et~al.}(2001){Kolokotronis}, {Basilakos}, {Plionis},
  \& {Georgantopoulos}}]{Kolokotronis_etal01}
{Kolokotronis}, V., {Basilakos}, S., {Plionis}, M., \& {Georgantopoulos}, I.
  2001, \mnras, 320, 49

\bibitem[{{Law} {et~al.}(2009){Law}, {Majewski}, \& {Johnston}}]{Law_etal09}
{Law}, D.~R., {Majewski}, S.~R., \& {Johnston}, K.~V. 2009, \apjl, 703, L67

\bibitem[{{Macci{\`o}} {et~al.}(2008){Macci{\`o}}, {Dutton}, \& {van den
  Bosch}}]{Maccio_etal08}
{Macci{\`o}}, A.~V., {Dutton}, A.~A., \& {van den Bosch}, F.~C. 2008, \mnras,
  391, 1940

\bibitem[{{Majewski} {et~al.}(2003){Majewski}, {Skrutskie}, {Weinberg}, \&
  {Ostheimer}}]{Majewski_etal03}
{Majewski}, S.~R., {Skrutskie}, M.~F., {Weinberg}, M.~D., \& {Ostheimer}, J.~C.
  2003, \apj, 599, 1082

\bibitem[{{Mandelbaum} {et~al.}(2006){Mandelbaum}, {Hirata}, {Broderick},
  {Seljak}, \& {Brinkmann}}]{Mandelbaum_etal06}
{Mandelbaum}, R., {Hirata}, C.~M., {Broderick}, T., {Seljak}, U., \&
  {Brinkmann}, J. 2006, \mnras, 370, 1008

\bibitem[{{Martin} {et~al.}(2008){Martin}, {de Jong}, \& {Rix}}]{Martin_etal08}
{Martin}, N.~F., {de Jong}, J.~T.~A., \& {Rix}, H. 2008, \apj, 684, 1075

\bibitem[{{Men{\'e}ndez-Delmestre} {et~al.}(2007){Men{\'e}ndez-Delmestre},
  {Sheth}, {Schinnerer}, {Jarrett}, \& {Scoville}}]{Menendez_etal07}
{Men{\'e}ndez-Delmestre}, K., {Sheth}, K., {Schinnerer}, E., {Jarrett}, T.~H.,
  \& {Scoville}, N.~Z. 2007, \apj, 657, 790

\bibitem[{{Merritt} \& {Quinlan}(1998)}]{Merritt_Quinlan98}
{Merritt}, D. \& {Quinlan}, G.~D. 1998, \apj, 498, 625

\bibitem[{{Moore} {et~al.}(2004){Moore}, {Kazantzidis}, {Diemand}, \&
  {Stadel}}]{Moore_etal04}
{Moore}, B., {Kazantzidis}, S., {Diemand}, J., \& {Stadel}, J. 2004, \mnras,
  354, 522

\bibitem[{{Navarro} {et~al.}(1996){Navarro}, {Frenk}, \&
  {White}}]{Navarro_etal96}
{Navarro}, J.~F., {Frenk}, C.~S., \& {White}, S.~D.~M. 1996, \apj, 462, 563

\bibitem[{{Neto} {et~al.}(2007)}]{Neto_etal07}
{Neto}, A.~F. {et~al.} 2007, \mnras, 381, 1450

\bibitem[{{Noordermeer} \& {Verheijen}(2007)}]{Noordermeer_Verheijen07}
{Noordermeer}, E. \& {Verheijen}, M.~A.~W. 2007, \mnras, 381, 1463

\bibitem[{{Olling} \& {Merrifield}(2000)}]{Olling_Merrifield00}
{Olling}, R.~P. \& {Merrifield}, M.~R. 2000, \mnras, 311, 361

\bibitem[{{Penarrubia} {et~al.}(2010){Penarrubia}, {Walker}, \&
  {Gilmore}}]{Penarrubia_etal10}
{Penarrubia}, J., {Walker}, M.~G., \& {Gilmore}, G. 2010, in the proceedings of
  "Hunting for the Dark: The Hidden Side of Galaxy Formation", Malta, 19-23
  Oct. 2009, eds. V.P. Debattista \& C.C. Popescu, AIP Conf. Ser.
  (astro-ph/1001.4500)

\bibitem[{{Sackett} \& {Sparke}(1990)}]{Sackett_Sparke90}
{Sackett}, P.~D. \& {Sparke}, L.~S. 1990, \apj, 361, 408

\bibitem[{{Simon} {et~al.}(2003){Simon}, {Bolatto}, {Leroy}, \&
  {Blitz}}]{Simon_etal03}
{Simon}, J.~D., {Bolatto}, A.~D., {Leroy}, A., \& {Blitz}, L. 2003, \apj, 596,
  957

\bibitem[{{Spekkens} \& {Sellwood}(2007)}]{Spekkens_Sellwood07}
{Spekkens}, K. \& {Sellwood}, J.~A. 2007, \apj, 664, 204

\bibitem[{{Springel} {et~al.}(2004){Springel}, {White}, \&
  {Hernquist}}]{Springel_etal04}
{Springel}, V., {White}, S.~D.~M., \& {Hernquist}, L. 2004, in IAU Symposium,
  Vol. 220, Dark Matter in Galaxies, ed. {S.~Ryder, D.~Pisano, M.~Walker, \&
  K.~Freeman}, 421--+

\bibitem[{{Stadel}(2001)}]{Stadel01}
{Stadel}, J.~G. 2001, Ph.D.~Thesis, Univ. of Washington

\bibitem[{{Tissera} {et~al.}(2010){Tissera}, {White}, {Pedrosa}, \&
  {Scannapieco}}]{Tissera_etal10}
{Tissera}, P.~B., {White}, S.~D.~M., {Pedrosa}, S., \& {Scannapieco}, C. 2010,
  \mnras, 406, 922

\bibitem[{{Trachternach} {et~al.}(2009){Trachternach}, {de Blok}, {McGaugh},
  {van der Hulst}, \& {Dettmar}}]{Trachternach_etal09}
{Trachternach}, C., {de Blok}, W.~J.~G., {McGaugh}, S.~S., {van der Hulst},
  J.~M., \& {Dettmar}, R. 2009, \aap, 505, 577

\bibitem[{{Valluri} {et~al.}(2010){Valluri}, {Debattista}, {Quinn}, \&
  {Moore}}]{Valluri_etal10}
{Valluri}, M., {Debattista}, V.~P., {Quinn}, T., \& {Moore}, B. 2010, \mnras,
  403, 525

\bibitem[{{Villalobos} {et~al.}(2009){Villalobos}, {Kazantzidis}, \&
  {Helmi}}]{Villalobos_etal10}
{Villalobos}, {\'A}., {Kazantzidis}, S., \& {Helmi}, A. 2010, ApJ, 718, 314

\bibitem[{{Zentner} {et~al.}(2005){Zentner}, {Kravtsov}, {Gnedin}, \&
  {Klypin}}]{Zentner_etal05}
{Zentner}, A.~R., {Kravtsov}, A.~V., {Gnedin}, O.~Y., \& {Klypin}, A.~A. 2005,
  \apj, 629, 219

\bibitem[{{Zwaan} {et~al.}(1995){Zwaan}, {van der Hulst}, {de Blok}, \&
  {McGaugh}}]{Zwaan_etal95}
{Zwaan}, M.~A., {van der Hulst}, J.~M., {de Blok}, W.~J.~G., \& {McGaugh},
  S.~S. 1995, \mnras, 273, L35

\end{thebibliography}

\end{document}